\documentclass[pre,twocolumn,showpacs]{revtex4}
\usepackage{graphicx}
\begin{document}
\title{Singular Scaling Functions in Clustering Phenomena}
\author{Mustansir Barma} 
\affiliation{Department of Theoretical Physics \\ Tata Institute of Fundamental Research \\ Homi Bhabha Road, Mumbai 400005, India}
\begin{abstract}
We study clustering in a stochastic system of particles sliding down a
fluctuating surface in one and two dimensions. In steady state, the
density-density correlation 
function is a scaling function of separation and system size.
This scaling function is
singular for small argument --- it exhibits a cusp singularity for
particles with mutual exclusion, and a divergence for noninteracting
particles. The steady state is characterized by giant fluctuations which do not
damp down in the thermodynamic limit.  The autocorrelation function is
a singular scaling function of time and system size.  The scaling
properties are surprisingly similar to those for particles moving in a
quenched disordered environment that results if the surface is frozen.
\end{abstract}
\maketitle
\section{Introduction}
\label{intro}

Clustering phenomena are widespread.  On the one hand, they include
the growth of regions of a homogeneous ordered phase, such as droplets
of a liquid or domains in a magnet; on the other hand, they also
include the formation of inhomogeneous structures such as those that
arise in granular gases.  In general, clusters break and re-form due
to the action of noise, whether thermal or induced by external
driving.  Concomitantly, the distribution of clusters -- their sizes
and spatial locations -- keep evolving in time.  A useful diagnostic
of the nature of clustering, both in steady state and during the
approach to it, is provided by the two-point density-density
correlation function.  In this paper, we will discuss the scaling
properties of correlation functions, and show that clustered states
of different sorts correspond to different types
of singularities in the corresponding scaling functions.

One of the most familiar and well studied examples of clustering is
afforded by phase ordering kinetics.  Here the clusters are simply
domains of ordered phases which form following a rapid quench from an
initially disordered state to a regime which favours ordering.  The
coarsening of ordered domains in time is captured by the two-point
correlation function; it shows simple scaling properties when spatial
separations are scaled by the mean domain size, itself a growing
function of time \cite{bray}.  In steady state, the system size
replaces the domain size in the scaling function.

There are a number of situations in which particles cluster
in such a way that the resulting density profile is patchy and far
from homogeneous.  For 
instance, particles may arrange themselves in a critical state with
power law decays of the correlation function, e.g. molecules in a
fluid at the critical point
\cite{critical} or galaxies in the universe \cite{pietronero}.
However, the sort of 
clustering of interest in this paper is different in that it is
associated with long range order, unlike a critical state, yet has
strong fluctuating properties which make it quite different from a
conventional phase ordering system.

We focus on particles subjected to randomly
fluctuating forces, which result in an inhomogeneously clustered
state.  Particles advected by fluid flows exhibit a clustering
tendency whose strength depends on the compressibility of the fluid
and the inertia of the particles \cite{maxey,balkovsky,gawedzki}. Earlier
studies have shown that in the absence of noise, particle trajectories
can coalesce as time advances \cite{deutsch,mehlig}.  Not
surprisingly, external noise --- 
which allows coalesced clusters to break up --- has a profound effect, 
and the dynamical steady state which results
from the balance between making and breaking clusters has interesting
characteristics. Our aim in this paper is to discuss the nature of
this state, with a focus on the scaling properties of correlation
functions. We study simple stochastic models of particles sliding down
a fluctuating surface
\cite{das1,das2,manoj,nagar1,nagar2,chatterjee}. As for
the case of phase ordering kinetics, the two-point
correlation function has simple scaling properties. However, now the
associated scaling function develops singularities -- these are {\em
cusp singularities} for particles with mutual exclusion, and {\em
divergences} for noninteracting particles.  The occurrence of such
singularities is quite robust, and survives as parameters of the model
and the dimension are changed.  Moreover, recent work has shown that
such singularities are found also in other completely different
systems which exhibit inhomogeneous clustering \cite{mishra,shinde}.

In a finite system, a dynamical steady state is reached in which
correlations are scaling functions of the separation and system size.
Like conventional phase ordering systems, this system exhibits long
range order, but unlike them, the scaling functions are singular
at small argument, indicative of a steady state which is intrinsically
different from a phase separated state in an equilibrium system.
A crucial point about the steady state is the existence of
strong fluctuations which do not die down in the thermodynamic limit,
due to which the degree of ordering in the system keeps fluctuating,
but never vanishes.

\section{Scaling in phase ordering systems}

Phase ordering kinetics deals with the onset of order
in a system which is brought rapidly from a disordered state to an
order-promoting regime \cite{bray}.  Domains of competing phases coarsen
in time, a process which is captured by the two-point correlation
function $C(r,t) = \langle \sigma_i(t) \sigma_{i+r}(t))\rangle$
where $\sigma_i = \pm 1$ is an Ising variable which describes a spin
in a magnet, or an occupancy variable $(1 + \sigma_i)/2$ within a
lattice gas description. In an infinite system, the
correlation function has the scaling form
\begin{equation}
C_S (r,t) = Y(r/{\cal L}(t))
\label{one}
\end{equation}
in the scaling limit $r \rightarrow \infty$, $t \rightarrow \infty$
with the ratio $y = r/{\cal L}(t)$ held fixed.  Here ${\cal L}(t)$ is
a growing, time-dependent length scale which describes the typical
linear size of a cluster at time $t$; typically it grows as a power
law in time, ${\cal L}(t) \sim t^{\beta'}$.


The scaling form of Eq. (1) holds only if the separation $r \gg$ the
correlation length $\xi$.  In general, the correlation function
$C(r,t)$ contains both the scaling part $C_S(r,t)$ and
an analytic part $C_A(r)$ which comes from short-distance $(r \ll \xi)$
correlations, independent of ${\cal L}(t)$. In practice, it is
important to account for the occurrence of $C_A(r)$ while analyzing
data for the correlation function to see whether scaling holds.

When the system size $L$ is infinite, the domain size ${\cal L}
(t)$ can grow indefinitely, with the scaling form of Eq. 1 continuing
to hold. However, if $L$ is finite, then Eq. 1 ceases to hold once $t$
is large enough that ${\cal L}(t) \approx L$. Beyond this, the system
reaches steady state and Eq. (1) is replaced by
\begin{equation}
\bar C_S (r,L) = \bar Y (r/L).
\end{equation}
There are two important characteristics of the scaling functions $Y(y)$
of Eq. 1 and $\bar Y(y)$ of Eq. 2: (i) the scaling
function has a finite intercept, (ii) the scaling function
falls linearly for small $y$.  Property (i) is important as the
intercept is a measure of long range order (LRO).  To see this for
$Y(y)$, recall that LRO is defined in an infinite system in steady state as
$m^2_0 = \lim_{r\rightarrow\infty} \langle \sigma_i \sigma_{i+r}\rangle$.  
Now consider two points separated by an arbitrarily large but
fixed distance $r$ in a coarsening system. As $t \rightarrow \infty$,
domain sizes grow without bound and the interior of each domain is
well approximated by steady state. Thus at large times, both points
would belong to the same domain with probability one, implying that their
correlation equals $m_0^2$. Since $t \rightarrow \infty$ leads to $y
\rightarrow 0$, no matter how large $r$ is, this implies that the
intercept in Fig. 1 is $m^2_0$.

\begin{figure}[h!]
\includegraphics[scale=0.4]{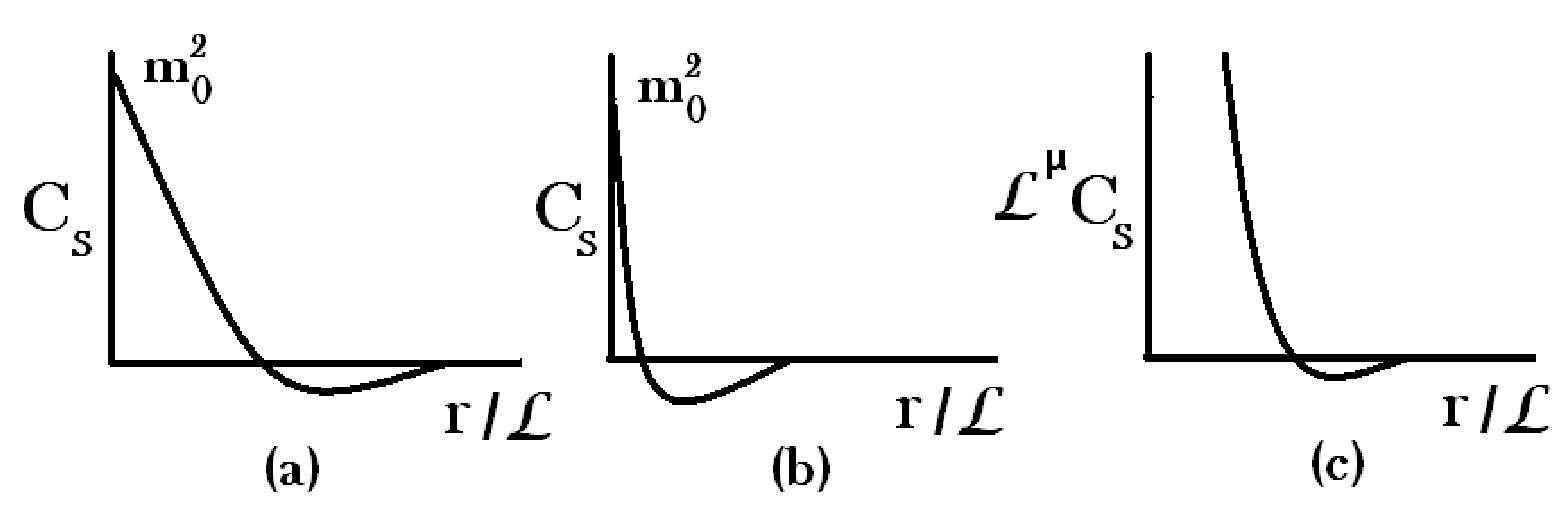}
\caption{The behaviour of the scaling part of the correlation function
is depicted for (a) normal phase ordering systems (b) a system of
particles with exclusion, sliding down a fluctuating surface (Model E)
(c) noninteracting particles sliding down a fluctuating surface (Model
F).  The 
linear drop of $C_S$ in (a) [the Porod law] is replaced by a
cusp singularity in (b), and by a divergence in (c).}
\end{figure}

Further, property (ii) which describes the manner in which the scaling
function falls from the intercept value $m_0^2$ is also
significant. For scalar order parameters, it falls linearly: $Y(y)
\approx m_0^2 - b|y|$ as $y \rightarrow 0$ \cite{bray}.  For a $d$-dimensional
system, this translates into a decay $\sim [k{\cal
L}(t)]^{-(d+1)}$ for the structure factor at large wave-vector
$k$. This power law decay is known as the Porod Law
\cite{porod}. The linear fall of correlations for $|y|=|r/ {\cal L}|
\ll 1$
can be traced to the existence of well-defined interfaces between two
phases: Two points a distance $r$ apart are positively correlated if
both belong to the same domain, and negatively correlated if a single
interface intersects the line joining them; for small values of
$r/{\cal L}$, the chance of this happening is proportional to $r$ and
inversely proportional to ${\cal L}$, accounting for the linear
drop of the scaling function at small argument.

In subsequent sections we investigate how far these properties are
respected when we have clustering of an inhomogeneous sort. We find
that there is a class of inhomogeneous clustering phenomena,
exemplified by models of particles sliding on randomly fluctuating
surfaces, for which most of the properties discussed above are valid,
except for the form of singular behaviour of the scaling function at
small argument.  These models and a discussion of the singularities
they display are discussed in the following sections.

\section{Model: particles sliding on a fluctuating surface}

The model consists of a system of overdamped particles sliding down
under gravity along the local slope of a fluctuating interface
(Fig. 2) \cite{das1,nagar1}.  In 1-d, we model the interface through the
single-step model \cite{liggett}.  It consists of a series of links,
with the slope of the link between the $i$'th and $(i+1)$'th site being
$\tau_{i+\frac{1}{2}} = \pm 1$, so that a
typical $\{\tau\}$ configuration is $/ / / \backslash
\backslash \backslash \backslash \backslash / \backslash /$.  The
height at site $i$ is 
given by $h_i = \sum^i_{j=1} \tau_{j-\frac{1}{2}}$. The elementary
dynamical move is a stochastic corner flip, which involves the
exchange of adjacent $\tau$'s: the transition $/ \backslash \rightarrow
\backslash /$ occurs with rate $p_1$, while the reverse transition
$\backslash / \rightarrow / \backslash$ occurs with rate $q_1$.  For
large $r$ 
and $t$, with $p_1 = q_1 = 1$ the interface dynamics reduces to the
Edwards-Wilkinson model \cite{edwards}, while with $p_1 \not= q_1$, it
reduces to the 
Kardar-Parisi-Zhang (KPZ) model \cite{kardar}.  Note that the overall
slope of the 
interface ${\cal T} = \sum^L_{i=1} \tau_{i+\frac{1}{2}}$, is
conserved; we use periodic boundary conditions and work with
interfaces with no overall tilt i.e. ${\cal T}=0$.

For the dynamics of the particles, there are two cases: particles with
mutual exclusion (E) \cite{das1}, and particles with no interactions, i.e. free
(F) \cite{nagar1}.  

\begin{figure}[h!]
\includegraphics[scale=0.4]{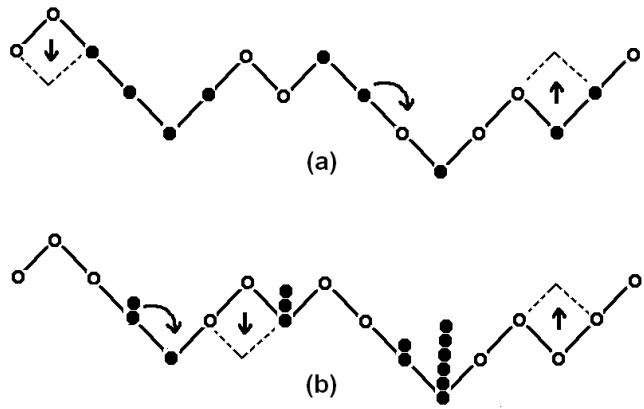}
\caption{Typical configurations and moves for lattice models of
particles sliding down a fluctuating surface in 1-$d$. (a) corresponds
to Model E (particles with exclusion) and (b) to Model F (particles
with no interactions).}
\end{figure}

In case E (Fig. 2a), a particle and hole at sites $i$ and $i+1$ exchange at a
rate which is governed by the slope variable $\tau_{i+\frac{1}{2}}$.
A move down the slope $(\bullet \backslash {\rm o} \rightarrow {\rm
o}\backslash\bullet$ or ${\rm o} / \bullet \rightarrow \bullet/{\rm o}$) occurs
with rate $p_2$, while the reverse transitions occur with rate $q_2 < p_2$.  

In case F (Fig. 2b), there is no limit on the occupancy of any site, and a
particle moves to its neighbouring site with rate $p_2$ or $q_2$
depending on whether it traverses the intervening $\tau$ bond in
the downward or upward direction.  If  $q_2=0$, particles move only downward and coalesce at the bottoms of valleys. Nevertheless, particle clusters can break again if they are carried, by surface evolution, to a local hilltop; subsequently, particles are equally likely to fall left or right, leading to a break-up of the cluster.

Notice that in both cases E and F, the particles are passive scalars: while the
evolution of particles occupancies $\{n_i\}$ is governed by the
interface configuration $\{\tau_{i+\frac{1}{2}}\}$, the evolution of
the interface is autonomous and is not influenced by the
particles.

Model E with EW dynamics is related to the Lahiri-Ramaswamy (LR) model
of sedimenting colloidal crystals \cite{lahiri1,lahiri2}.  In the
present context, the general LR model allows for the particles to act
back on the fluctuating surface, in which case phase-separated and
disordered phases ensue \cite{lahiri2,das3}.  The case of
passive particles under consideration lies at the boundary
between these more conventional phases \cite{ramaswamy}.

A lattice model closely related to model F with KPZ dynamics was
studied by Drossel and Kadar \cite{drossel1,drossel2}; their results
are discussed in Section 5 along with ours.  
Interestingly, model F with KPZ dynamics relates to the problem of
passive scalar 
advection by a noisy Burgers flow.  This follows from the well known
mapping \cite{kardar} between the KPZ equation and the noisy Burgers
equation which describes a compressible flow, and from the observation
that movement of a particle along the local slope in the present
context translates, through
the mapping, to a particle being advected by the local velocity
field.

\section{Sliding particles with exclusion}

We turn to the properties of a system of particles with mutual
exclusion on a fluctuating interface, evolving under the dynamics  
discussed in the previous section. 

\subsection{Coarsening}

At $t=0$, we start out with a configuration of the interface drawn
from steady state, and particles distributed randomly on sites.
Numerical simulations of this process using a very large system size
were performed to monitor the two-point correlation functions
$C(r,t)$, averaged over many histories \cite{das1,das2}.  Particles
and surface elements were updated at equal rates.  The results show
that there is a good scaling collapse of the data to the form of Eq. 1, with
${\cal L} (t) \sim t^{1/z}$ where $z$ is the dynamical exponent which
characterizes the interface dynamics.  This is because in time $t$,
new valleys of size $t^{1/z}$ form, so that particles in basins of
that size are drawn together.  In 1 dimension, we have $z = 2$ for EW
evolution, and $z = 3/2$ for KPZ evolution.  

The significant new point is that the scaling function has a cusp singularity
\begin{equation}
C_S (y) \approx m^2_0 - b|y|^\alpha ~~~ {\rm as}~ y \rightarrow 0
\end{equation}
with $y = r/{\cal L} (t)$.  This translates into a tail $\sim (k{\cal
  L}(t))^{1+\alpha}$ for the wave-vector dependence of the structure
  factor, implying non-Porod behaviour.  The cusp exponent $\alpha$ is found to be $\simeq 0.5$
  with EW evolution and $\simeq 0.25$ with KPZ evolution of the
  interface \cite{das2}.  The value of $\alpha$ is found to be unaffected by
  variations of the density of particles or the relative
  ratio of interface and particle update rates.  Moreover, the
  phenomenon persists with 2-d interfaces as well \cite{manoj}; with
  KPZ dynamics, 
  it is found that $\alpha \simeq 0.38$ (see Section 4.4).

\subsection{Steady state and Cluster size distribution}

A study of finite systems of size $L$ shows that in steady state, the
correlation function is a scaling function of separation and size, as
in Eq. 2 \cite{das2}.  Once again, the scaling function shows a cusp
  singularity, 
and the value of the cusp exponent $\alpha$ matches that obtained in
the coarsening studies discussed in Section 4.1.

A clue to the nature of the state comes from a study of the
distribution $P(\ell)$ of cluster sizes $\ell$ \cite{das2}.  A cluster
is defined 
as a stretch of continuously occupied sites, with unoccupied perimeter
sites.  In the 1-d EW case, it is found that $P(\ell)\sim
\ell^{-\theta}$ with $\theta \simeq 1.8$; symmetry implies the
same form for hole clusters.  In the 1-d KPZ case, simulations reveal that
hole clusters follow a power law decay with $\theta \simeq 1.85$,
whereas the decay for particle clusters do not seem to follow a simple
power law.

The power-law decay of $P(\ell)$ can be related to the cusp observed in
the two-point correlation function \cite{das2} within the independent interval
approximation (IIA) which is based on the premise that the occurrence
of successive intervals (clusters of particles and holes) are
independent events.  With this approximation, the Laplace transforms
$\tilde P(s) = \int^\infty_0 d\ell~ e^{-\ell s} P(\ell)$ and $\tilde
C(s) = \int^\infty_0 dr~ C(r)$ are related by \cite{sire}
\begin{equation}
s\left[1-s\tilde C(s)\right] = \frac{2}{<\ell>} \frac{1-\tilde
  P(s)}{1+\tilde P(s)}
\end{equation}
where $<\ell>$ is the mean cluster size.  For the slow power law decay
$P(\ell) \sim \ell^{-\theta}$, where $\ell$ is limited by $\ell = L$,
we have $<\ell> \approx a L^{2-\theta}$ and $\tilde P(s) \approx 1 -
bs^{\theta -1}$ for $1/L \ll s \ll 1$.  Equation 4 then leads to the
scaling from of Eqs. 2 and 3, with the cusp exponent given by 
\begin{equation}
\alpha = 2-\theta.
\end{equation}

The occurrence of LRO in this fluctuating system can also be understood by
noting that extremal statistics applied to $P(\ell) \sim
\ell^{-\theta}$ with $1 < \theta < 2$ leads to the conclusion that the
largest clusters are of size $\sim L$. 

\subsection{Adiabatic limit: Coarse-grained depth (CD) model}

We can calculate $\theta$ and thus $\alpha$ within the IIA for a
class of models motivated by considering the adiabatic limit
of infinitely slow interface evolution. Then the height field
$\{h_i\}$ is a quenched random variable with $h_i = \sum^i_{j=1}
\tau_{j-\frac{1}{2}}$.  The steady state of the particle distribution
is a thermal equilibrium state with temperature defined by $T = V \ell
n (p_2/q_2)$ where $V$ is the potential energy drop across a single
link.  In the limit of low $T$, this equilibrium state is
characterized by a filling level akin to the Fermi level, namely the
height up to which the particles are filled. In this limit, the
correlation function $<{s_i s_{i+r}}>$ 
coincides with $C(r)$.  

This motivates us to define and study more general coarse-gained depth
models (CD models) \cite{das1} with variable filling levels 
\begin{equation}
s_i(t) = sgn\left[h_i(t) - \bar h\right]
\end{equation}
where $\bar h$ is the height of a surface cut.  Another natural choice
for $\bar h$ is the instantaneous average height of the configuration
$\{h_i\}$, but with this definition the cut need not coincide with the
filling level defined in the previous paragraph.  Yet another
possibility for the cut is to pin it to the instantaneous height of
the first site.  For the 1-d EW and KPZ interfaces, the configuration
of the interface is given by the trajectory of a random walk and the
filling level defines a cut of this trajectory.  The intervals between
successive crossings then follow a distribution $p(\ell) \sim
\ell^{-3/2}$. Using Eq. 5, this leads to a cusp exponent $\alpha =
1/2$.

\subsection{Higher dimensions}

Singularities in scaling functions are also found to occur in a system
of particles sliding down a fluctuating rough surface in two dimensions
\cite{manoj}.  The surface is modeled by a discrete solid-on-solid model with a
simple growth rule: a site on a square lattice is selected a random and
its height is decreased by 2 units, provided the height at all four 
of its neighbouring sites is lower.  The asymptotic properties of this
model are expected to be the same as those of the (2+1)-dimensional
KPZ equation.  Particles are initially distributed at random, and
updated by choosing a random particle, attempting to move it to a
randomly chosen neighbour, and actually moving it provided the local
slope is downward and the target site is unoccupied.  

Results on coarsening \cite{manoj} indicate that Eq. 1 holds, with
${\cal L}(t) \sim t^{1/z}$ and $z \simeq 1.6$, close to the value of
the dynamical exponent which defines the relaxation of a
(2+1)-dimensional KPZ surface.  The cusp exponent $\alpha$
characterizing the decay of the scaling form of the density-density
correlation function is $\simeq 0.38$.  A study of the CD
model (with $\bar h$ taken to be the average height) shows that Eq. 2
holds, with $\alpha \simeq 0.43$.

In steady state, we may relate the cusp exponent $\alpha$ in the CD
model to first return probabilities through the following argument
\cite{manoj}.  Consider a self-affine surface with roughness exponent
$\chi$, and define $P(\ell)$ as the probability that the surface first
returns to its starting height a distance $\ell$ along an arbitrary
linear direction.  $P(\ell)$ decays as power law for $\ell \ll L:
P(\ell) \sim \ell^{-(2-\chi)}$.  On applying the IIA to different
segments on a straight line connecting two points in a distance $r$
apart, Eq. 5 leads to the conclusion that $\alpha = \chi$, i.e. the
cusp exponent for the CD model is determined by the roughness exponent
of the fluctuating, driving surface.  This is borne out by the results
in both 1 and 2 dimensions ($\alpha = \chi = 1/2$ in 1-d; $\alpha
\simeq 0.43, \chi \simeq 0.4$ in 2-d).

\subsection{Dynamics}

As with static (equal-time) correlation functions in steady state,
dynamical properties also exhibit scaling characterized by singular
scaling functions \cite{chatterjee}.  The autocorrelation function
$A(t) = \langle \sigma_i(0) \sigma_i(t)\rangle$  was monitored by Monte
Carlo simulation, and found to be described by a scaling function
$W(t/L^{z})$, in the scaling limit $t \rightarrow \infty, L \rightarrow
\infty$ with $t/L^{z}$ held finite. Parallel to the static situation, the scaling function $W_S(w)$ exhibits a cusp singularity at small argument:
\begin{equation}
W_S(w) \approx m_{0}^2(1-b'|w|^{\beta�}) ~~~~~~~ w \rightarrow 0
\end{equation}
where $w = t/L^z,m_0^2$ measures long-range order and $\beta�$ is the cusp
exponent. The results of simulations indicate that $\beta� \simeq
0.22$ for Edwards-Wilkinson surface dynamics and $\beta� \simeq 0.18$
for KPZ surface dynamics \cite{chatterjee}.

Just as for static correlations (section 4.3), insight
into singular scaling for dynamic correlations can be obtained from 
analytic calculations on the corresponding CD model. To
this end, the autocorrelation function $ \langle s_i(0) s_i(t)
\rangle$ of the CD model with EW evolution was calculated
\cite{chatterjee} and shown
to follow the form of Eq. 2 with $m_0=1$ and $\beta� = 0.25$.

Further, aging functions of the form $\mathcal {A} (t_{1},t_{2})
\equiv \langle \sigma_i(t_1)\sigma_i(t_2)\rangle$
were investigated for a coarsening system, starting from 
a state with randomly distributed particles \cite{chatterjee}.  If 
$t_1,t_2 \gg 1$, the 
aging function ${\cal A}$ is a function of the ratio $t_1/t_2$
alone. For $t_1 \gg t_2$, we find ${\cal A } \approx
m_0^2[1-b_1(t_2/t_1)^{\beta}]$, whereas for $t_1 \ll t_2$, the function
${\cal A }$ follows a power law decay
$(t_1/t_2)^{\gamma}$. Simulations show that $\gamma \simeq 0.69$ for
EW interface dynamics, $\gamma \simeq 0.82$ for KPZ dynamics, and 
$\gamma \simeq 0.82$ for the CD model.

\subsection {Ordering and Giant fluctuations}
 
As we have seen, the two-point correlation function characterizing the
steady state approached by a system of particles driven by a
fluctuating surface exhibits long range order. Let us ask for a {\it
  one-point} function or order parameter which can be used to
characterize the steady state.  To this end, we monitored the Fourier
transform 
\begin{equation}
Q(k) = \left|{1 \over L} \sum^L_{j=1} e^{ikj} n_j\right|, \ \ k =
{2\pi m \over L},
\end{equation}
where $m=1,2,\cdots,L-1$. The expectation value $Q_1^*=\langle Q_1 \rangle$
of the longest-wavelength mode $Q_1 = Q(2\pi/L)$ corresponding to
$m=1$ is a putative order parameter. For a state which exhibits
complete phase separation in a half-filled system, $Q_1^* \simeq
0.32$, while for a disordered state, $Q_1^*=0$; however, as we will 
see below, $Q_1^*=0$ does not necessarily imply a disordered
state. Figure 3 shows schematically the behaviour of $\langle Q(k)
\rangle$ as a function of $k$ for various values of $L$. For fixed
wave-vector $k \ne 0$, we see that $\langle Q(k) \rangle \rightarrow
0$ as $L \rightarrow \infty$. To study the $k \rightarrow 0$ limit, we
fix the mode number $m$ (e.g. $m=1$ or $m=2$) and monitor $ Q_m^* =
\langle Q_m \rangle \equiv \langle Q(2\pi m/L) \rangle$. We see that
$Q_1^*$, $Q_2^*$ etc. approach a finite limit as $L \rightarrow
\infty$. Numerical simulations reveal that for particles sliding down
a fluctuating EW surface, $Q_1^*\simeq 0.18$ and $Q_2^*\simeq 0.09$,
while $Q_1^*\simeq 0.16$ and $Q_2^* \simeq 0.08$ for the case of driving by
a KPZ surface.

A crucial point about the low-$m$ Fourier modes $Q_{m}$ is that they
have broad distributions in the thermodynamic limit. In time, the
state shows strong fluctuations in the type and amount of ordering,
within a subspace of ordered states. This leads us to call this
phenomenon fluctuation-dominated phase ordering (FDPO).  It is quite
unlike normal ordering in equilibrium phase transitions, where the
order parameters, such as the density in a liquid-vapour system or the
magnetization in a ferromagnet, have well-defined values at all times in large
systems. The order parameter distributions $Prob(\rho)$
and $Prob(m)$ in these cases consist of two delta functions at values
characteristic of each phase, i.e. at well-defined densities $\rho_v$
and $\rho_{l}$ for the fluid, or at magnetizations $\pm m_{0}$ for the
magnet. By contrast, in our system, $Prob(Q_1), Prob(Q_2)$
etc. approach well-defined {\it broad} distributions in the thermodynamic
limit.

\begin{figure}[h!]
\includegraphics[scale=0.4]{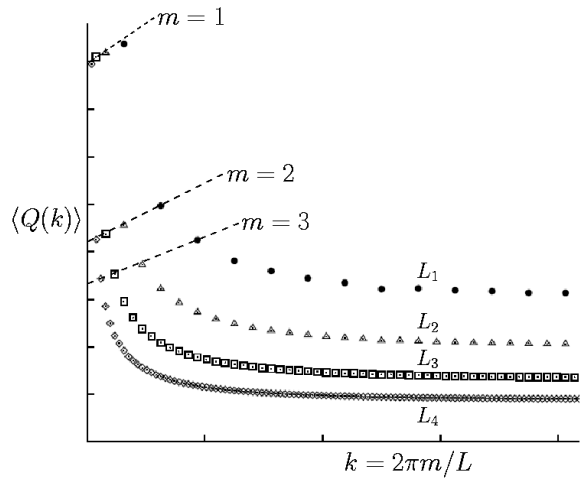}
\caption{Plots of $\langle Q(k)\rangle$ versus $k$ are shown for
various lattice sizes $L_1 < L_2 < L_3 < L_4$.  As $L$ increases, 
for each fixed low value
of $m$, $Q^\star_m = \langle Q_m\rangle$ approaches a limiting value
whose significance is discussed in the text.}
\end{figure}

The values of $Q_{1}^{*}, Q_{2}^{*}, \cdots$ characterize the sort of
clustering 
that occurs in the system. For instance, $Q_{1}$ is largest in
configurations with a single dense cluster of particles which extends
across half the system. On the other hand, $Q_{2}$ is largest in
configurations with two separated dense clusters. The time series for
$Q_1$ shows that it makes large excursions around its mean value,
occasionally reaching high values $\simeq 0.32$, and occasionally low
values $\simeq 0$ \cite{das2}. It is observed that a dip in the value of $Q_{1}$ is
usually accompanied by a rise in the value of $Q_{2}$. On the rare
occasions when $Q_{1}$ and $Q_{2}$ are both small, $Q_{3}$ picks up. These
observations are consistent with strong fluctuations of the ordered
structure: a single high-density region is most likely, but it
occasionally breaks into two, and more infrequently into three such
regions, and so on --- never, however, lapsing into a disordered state.
FDPO thus involves the system circulating within an attractor of states,
each with macroscopic order. In view of the strong anticorrelations
between $Q_{1}, Q_{2}, Q_{3}, \cdots$, the best characterization of
the state would be through the joint probability distribution
$Prob(Q_{1},Q_{2},\cdots)$ but this remains to be studied
systematically. 

Evidence for this picture comes from monitoring the length
$\ell_{max}(t)$ of the longest cluster in a system of size $L$
\cite{chatterjee}.  In a disordered state, $\langle
\ell_{max}\rangle$ scales as $\log L$.  A numerical study of the
probability distribution of $\ell_{max}$ shows that $\langle
\ell_{max}\rangle$ increases as $L^\phi$ where the exponent $\phi$
ranges from 0.6 to 0.9 for different sorts of surface dynamics, and
the full distribution scales with $\langle \ell_{max}\rangle$,
implying that the system does not reach a disordered state.   

\section{Sliding noninteracting particles}

The related problem of noninteracting particles advected by the
Kuramoto-Shivashinsky equation was 
studied by Bohr and Pikovsky \cite{bohr}, while Chin \cite{chin}
studied advection by a KPZ field.  In both these studies, no noise
acts on the particles, in which case clusters do not break up.
Besides studying cluster coalescence in time, they monitored the rms
displacement of a tagged particle, and found that it grows as
$t^{1/z}$ where $z = 3/2$ is the dynamic exponent of the 1-$d$ KPZ 
model.  Drossel and Kardar \cite{drossel2} studied the rms
displacement in the presence of noise, and found the same behaviour,
while Gopalakrishnan \cite{gopal} studied the same quantity for
driving by an EW surface.  Further, Drossel and Kadar 
studied the density-density correlation function in the cases of
particles falling along 
(advection) and against (anti-advection) the local slope. In
the latter case they concluded \cite{drossel1} that correlations show a
power law decay $(\sim r^{- \lambda})$ with increasing separation
$r$. We will discuss this further below in the light of our results.

\subsection{Correlation functions and number distribution}

Using Model F defined in Section 3 (see Fig. 2b) numerical simulations
were performed to monitor the steady state correlation function
$C(r,L) = \langle n_i n_{i+r}\rangle_L$ \cite{nagar1,nagar2}.  Both EW
and KPZ surfaces were 
considered in 1-$d$. Since there is no limit 
on the particle occupancy $n_i$ at any site, the possibility arises of a
much larger degree of clustering.  This manifests itself in the
fact that the correlation function is a scaling function of $r$ and
$L$. The small-argument singularity of the scaling function shows  
a divergence: 
\begin{equation}
\bar C_S(r,L) = L^{-\mu} \bar Y\left({r \over L}\right)
\end{equation} 
with
\begin{equation}
\bar Y(y) \sim y^{-\nu} \ \ \ {\rm as} \ y \rightarrow 0
\end{equation}
The divergence is strongest for the case of KPZ advection, in which
case $\nu =3/2$ and $\mu = 1/2$.  This result agrees with the
analytical results of Derrida et al. \cite{derrida} for a slightly
different model, 
which consists of two particles which slide down slopes, but are not
passive in that they block evolution on the sites at which they
reside. In Section 5 below we will see that this form of the
correlation function is found also in the adiabatic limit of a
quenched interface.

In the case of KPZ anti-advection, simulations show that Eqs. 9 and
10 hold, with $\mu = 0$ and $\nu \simeq 0.31$. The important
point is that $C$ is a function of $r/L$ and not 
$r$ alone. Thus the power law $\sim r^{-\nu}$ has an $L$-dependent prefactor,
unlike the power law behaviour at a critical point. In this
respect, our result differs from that of \cite{drossel1,drossel2}.

In the case of EW evolution,  Eqs. 9 and 10 are found to hold
again, with $\mu = 0$ and $\nu \simeq 2/3$. 

The probability that a given site holds $n$ particles $P(n,L)$ also
assumes the scaling form
\begin{equation}
P(n,L) \sim L^{-2\delta} f\left({n \over L^\delta}\right)
\end{equation}  
where 
\begin{equation}
f(y) \sim y^{-\gamma} \ \ \ {\rm as} \ y \rightarrow 0.
\end{equation}
Numerical simulations
for the case of KPZ advection yield $\delta \simeq 1$, $\gamma \simeq
1.15$; the numerical results are also consistent with $\gamma = 1$
with logarithmic corrections. For the case of KPZ anti-advection, we
find $\delta \simeq 1/3$, $\gamma \simeq 1.7$ while for EW surface dynamics we 
find $\delta \simeq 0.68$, $\gamma \simeq 1.5$.

Similar results were obtained in two dimensions, using a lattice model
similar to that discussed in Section 4.4.  The two point correlation
function shows a divergence $\sim (r/L)^{-\nu_2}$ with $\nu_2 \simeq
1.4$ for KPZ advection, $\nu_2 \simeq 0.5$ for KPZ antiadvection and
$\nu_2 \simeq 0.3$ for EW dynamics.

\subsection {Adiabatic limit: the Sinai model}

Consider the adiabatic limit in which the surface is absolutely still,
while particles evolve by performing biased random walks, with the
bias on each link set by the slope of the quenched surface.  The
problem then reduces to the Sinai problem of noninteracting random walkers
moving in a random medium which itself is defined as the trajectory of
a random walk. Since the surface is static, the walkers have infinite
time to explore the landscape, and eventually reach an equilibrium
state defined by a temperature $T= V \ln(p_2/q_2)$.  Thus $\langle n_i
n_{i+r}\rangle$ is given by $\exp[-\beta(h_i + h_{i+r})]/Z(\{h_k\})$ for
a particular configuration of heights $\{h_k\}$.  We need to further 
average $\langle n_i n_{i+r}\rangle$ over quenched disordered
landscapes $\{h_k\}$, a computation 
that was carried out by Comtet and Texier \cite{comtet}.  Taking the
scaling limit $r \rightarrow \infty$, $L \rightarrow \infty$ with
$r/L$ held fixed, we find that $\overline{\langle n_i n_{i+r}\rangle}$
has the scaling form of Eq. 2, with 
\begin{equation}
\bar Y(y) = (2n \beta^2 L)^{-1/2} [y(1-y)]^{-3/2}
\end{equation}
where $y = r/L$.  

Further, the probability $P(n,L)$ that a site holds $n$ particles can
be calculated as well \cite{chatterjee}.  It has the scaling form
$P(n,L) = 4/\beta^4 L^2 \bar X(x)$ with $x = 2n/\beta^2 L$ and
\begin{equation}
\bar X(x) = e^{-x} K_0 (x)/x
\end{equation}
where $K_0 (x)$ is a Bessel function of imaginary argument.

Surprisingly, we find very good agreement between these results for
an equilibrium state of particles in a disordered environment, and
those for the strongly nonequilibrium situation of KPZ advection
discussed in Section 5.1.  The correlation scaling function $\bar
Y(y)$ in Eq. 13 fits the numerical data closely, if we set $\beta
\simeq 4$.  On the other hand, the scaling function $\bar X(x)$ in
Eq. 14 
describes the probability density data for the nonequilibrium system
provided we set $\beta \simeq 2.3$.  Although the driving
force is different in the two cases -- being the 
temperature in the equilibrium disordered system, and surface
fluctuations in the nonequilibrium system -- it allows particles to
explore the terrain in both cases, and reach states with similar
though not identical characteristics.
  
\section{Conclusion}

The principal conclusion of this paper is that in a class of
inhomogeneously clustered states, the density-density correlation
function is a function not of the separation alone, but rather the ratio
of separation to system size (in steady state) or separation to a
growing length scale (in the coarsening regime).  This scaling
behaviour is reminiscent of phase ordered states, but unlike conventional
phase ordering, the scaling function here is singular at small
argument.  In a system of particles with mutual exclusion sliding down
a fluctuating surface, the singularity in question is a cusp, while
for noninteracting particles which can cluster more strongly, the
singularity is a divergence.  There is a degree of universality in the
exponents which characterize the singularities in scaling functions:
they remain unchanged under variation of parameters in the model, for
instance, the ratio of rates of particle movement and surface
fluctuations, though they do depend on the symmetry and
dimension of the driving field.

The other hallmark of the states under consideration here is the
occurrence of giant fluctuations, characterized by the standard
deviation of the order parameter growing proportionally to the mean,
as a consequence of a probability distribution that remains broad in
the thermodynamic limit. 

It is to be emphasized that the fluctuation-dominated states under
discussion here are quite 
different from critical states. In critical states, correlation
functions decay as power-laws, and size effects appear as
corrections.  By contrast, correlation functions in our case are, even
in leading order, functions of the separation scaled by system size.
Moreover, fluctuations at critical points lead to probability
distributions that narrow down in the thermodynamic limit, unlike the
giant fluctuations present here.

The occurrence of singular scaling functions is not confined to the
models studied in this paper.  A study of nonequilibrium nematic
states shows that the density-density correlation function shows
similar scaling properties, with a cuspy behaviour of the scaling
function \cite{mishra}.  Further, a recent study of inelastically colliding
particles for which the coefficient of restitution depends on the
relative velocity of approach of two particles, shows similar
non-Porod scaling properties \cite{shinde}.

Finally, there is an intriguing connection between the properties of
the strongly nonequilibrium system under study and an equilibrium
system with quenched disorder which is obtained by freezing the
fluctuating potential.  We have presented evidence for the
similarities of several properties, but a deeper understanding of the
connection remains an open question.

\noindent {\it Acknowledgements}: I would like to thank Dibyendu Das,
Manoj Gopalakrishnan, Apoorva Nagar, Sakuntala Chatterjee and Satya
Majumdar for enjoyable collaborations on the passive particle problems
discussed in this paper.  I am grateful to Shamik Gupta for useful
comments on the manuscript.


\begin{thebibliography}{}
\bibitem{bray} A.J. Bray, Adv. Phys. \textbf{43} (1997) 357.
\bibitem{critical} H.E. Stanley,  {\it Introduction to Phase Transitions
and Critical Phenomena} (Oxford, New York and Oxford, 1971).
\bibitem{pietronero} P.H. Coleman and L. Pietronero,
  Phys. Rep. \textbf{213} (1992) 311.
\bibitem{maxey} M. Maxey, J. Fluid Mech. \textbf{174} (1987) 441.
\bibitem{balkovsky} E. Balkovsky, G. Falkovich and A. Fouxon,
Phys. Rev. Lett. \textbf{86} (2001) 2790.
\bibitem{gawedzki} K. Gawedzki and M. Vergassola, Physica D
\textbf{138} (2000) 63.
\bibitem{deutsch} J.M. Deutsch, J. Phys. A\textbf{18} (1985) 1449.
\bibitem{mehlig} B. Mehlig and M. Wilkinson,
Phys. Rev. Lett. \textbf{92} (2004) 250602. 
\bibitem{das1} D. Das and M. Barma, Phys. Rev. Lett. \textbf{85}
(2000) 1602.
\bibitem{das2} D. Das, M. Barma and S.N. Majumdar,
Phys. Rev. E\textbf{64} (2001) 046126.
\bibitem{manoj} G. Manoj and M. Barma, J. Stat. Phys. \textbf{110}
(2003) 1305.
\bibitem{nagar1} A. Nagar, M. Barma and S.N. Majumdar,
Phys. Rev. Lett. \textbf{94} (2005) 240601.
\bibitem{nagar2} A. Nagar, S.N. Majumdar and M. Barma,
Phys. Rev. E\textbf{74} (2006) 021124.
\bibitem{chatterjee} S. Chatterjee and M. Barma,
Phys. Rev. E\textbf{73} (2006) 011107. 
\bibitem{mishra} S. Mishra and S. Ramaswamy,
Phys. Rev. Lett. \textbf{97} (2006) 090602.
\bibitem{shinde} M. Shinde, D. Das and R. Rajesh, arXiv condmat
0708.0930 (2007).
\bibitem{porod} G. Porod, in {\it Small Angle $X$-ray Scattering},
edited by O. Glatter and L. Kratky (Academic Press, New York, 1983).
\bibitem{liggett} T.M. Liggett, {\it Interacting Particle Systems},
Springer (2000).
\bibitem{edwards} S.F. Edwards and D.R. Wilkinson,
Proc. R. Soc. London, Ser. A\textbf{381} (1982) 17.
\bibitem{kardar} M. Kardar, G. Parisi and Y.C. Zhang,
Phys. Rev. Lett. \textbf{62} (1986) 89.
\bibitem{lahiri1} R. Lahiri and S. Ramaswamy,
Phys. Rev. Lett. \textbf{79} (1997) 1150.
\bibitem{lahiri2} R. Lahiri, M. Barma and S. Ramaswamy,
Phys. Rev. E\textbf{61} (2000) 1648.
\bibitem{das3} D. Das, A. Basu, M. Barma and S. Ramaswamy,
Phys. Rev. E\textbf{64} (2001) 021402.
\bibitem{ramaswamy} S. Ramaswamy, M. Barma, D. Das and A. Basu,
Phase Transit. \textbf{75} (2002) 363.
\bibitem{drossel1} B. Drossel and M. Kardar,
Phys. Rev. Lett. \textbf{85} (2000) 614.
\bibitem{drossel2} B. Drossel and M. Kardar, Phys. Rev. B\textbf{66}
(2002) 195414.
\bibitem{bohr} T. Bohr and A. Pikovsky, Phys. Rev. Lett. \textbf{70}
(1993) 2892.
\bibitem{chin} C.S. Chin, Phys. Rev. E\textbf{66} (2002) 021104.
\bibitem{gopal} M. Gopalakrishnan, Phys. Rev. E\textbf{69} (2003)
011105.
\bibitem{sire} S.N. Majumdar, C. Sire, A.J. Bray and S.J. Cornell,
Phys. Rev. Lett. \textbf{77} (1996) 2867.
\bibitem{derrida} B. Derrida, S.A. Janowsky, J.L. Lebowitz and
  E.R. Speer, J. Stat. Phys. \textbf{73} (1993) 813.
\bibitem{comtet} A. Comtet and C. Texier, {\it Supersymmetry and
Integrable Models Proceedings, Chicago, IL}, edited by H. Aratyn,
T. Imbo, W.Y. Keung and U. Sukhatme (Springer, Berlin, 1998).

\end{thebibliography}
\end{document}